\documentclass[11pt,twocolumn]{article}
\usepackage{epsfig}
\usepackage{latexsym}
\usepackage{amssymb}
\usepackage{times}

\makeatletter
\def\section{\@startsection{section}{1}{\z@}{-3.5ex plus-.3ex minus-.2ex}%
{1.5ex plus.3ex}{\small\bfseries}}
\renewcommand{\@makecaption}[2]{\footnotesize Fig.\hspace{-.2em}#1.\ \ #2}
\makeatother

\begin{document}
\twocolumn[\vspace*{1.7cm}
\title{Floating of critical states and the QH to insulator transition}
\author{\vspace{.2cm}{\large H. Potempa and L. Schweitzer\footnotemark[1]}  
\\[1.7\baselineskip]
{\footnotesize\itshape Physikalisch-Technische Bundesanstalt, Bundesallee 100,
38116 Braunschweig, Germany}}
\date{}
\maketitle
\vspace{-.5cm}
\rule{\textwidth}{.5pt}

\vspace{-.35cm}
\small
\begin{abstract}
The transition from the quantum Hall state to the insulator is considered 
for non-interacting electrons in a two-dimensional disordered lattice
model with perpendicular magnetic field. 
Using \textit{correlated} random disorder potentials the floating up
of the critical states can be observed in a similar way as in the 
continuum model. 
Thus, the peculiar behaviour of the lattice models reported previously 
originates in the special choice of uncorrelated random disorder
potentials.
\end{abstract}

\vspace{.25cm}

\rule{\textwidth}{.5pt}
\vspace{.5cm}
]
\footnotetext[1]{Corresponding author. Fax: +49\,531\,5928106; 
email: Ludwig.Schweitzer@ptb.de.}

\normalsize
\section{Introduction}
Instead of floating up in energy, as has been proposed \cite{Lau84,Khm84} 
and shown to exist in continuum models \cite{And84,KHA95}, a peculiar 
annihilation of the current carrying states takes place in lattice
models when the quantum Hall to insulator transition is approached. 
According to recent investigations 
\cite{LXN96,SW98,PBS98,YB99,HIM99}, increasing the disorder causes 
states associated with a negative Chern number to start moving outwards 
from the band centre and to destroy states with corresponding
positive Chern numbers at the band edges. 
This behaviour is a special feature of lattice
models which implicitly contain a periodic potential, but is not possible 
in the continuum model without a periodic potential usually considered 
to explain the levitation of the current carrying states.

The transition from the quantum Hall phase to the insulator is strongly 
affected by the way the critical states disappear. 
While in the continuum model the crossover takes place when the last 
current carrying state floats across the Fermi level, 
a direct transition to the insulator is possible in the lattice model 
also from higher Hall plateaus. The latter possibility is not contained 
in the global phase diagram for the quantum Hall effect \cite{KLZ92}
since it was based on the levitation scenario \cite{Lau84,Khm84}.
  
Up to now the results of experimental investigations are not conclusive 
\cite{SKD93,GJJ95,KMFP95,Hea99c}. Both, direct transitions from 
higher Hall plateaus to the insulator have been reported as well as
experiments that are in accord with the global phase diagram. 
Recently, it has been pointed out that the low field quantum Hall to
insulator transition might be unaccessible for the present time
\cite{Huc00}, because the available temperature range and sample sizes  
refuse to enter the scaling regime. However, a QH to insulator
transition as a function of disorder strength at higher magnetic
fields is more likely within experimental reach. 

In the present paper we consider the possible origin of both the diverging
experimental and theoretical results reported so far. In particular, 
the fate of the current carrying states is studied when the disorder is
increased. We also investigate, how the QH to insulator
transition depends on the correlation length of the disorder
potentials.

\section{Model with correlated disorder potentials}
A two-dimensional tight binding model describing non-interacting electrons 
moving in a perpendicular magnetic field and random disorder potentials 
associated with the lattice sites $k$ is given by the Hamiltonian
\begin{equation}
H=\sum_{k} w_k |k\rangle\langle k| + 
\sum_{<k \ne l>} V_{kl} |k\rangle\langle l|.
\label{AH}
\end{equation} 
The transfer terms $V_{kl}$ connect only neighbouring sites and
contain the magnetic field $B=\alpha h/e a^{-2}$ 
which is taken to be commensurate with the lattice. It is expressed as the 
number $\alpha=p/q$ of flux quanta $h/e$ per plaquette of area $a^2$ and 
enters via the complex phases, $V_{kl}\sim\exp(i2\pi \alpha k_x/a)$. 
We take $p/q=1/8$ so that the tight binding band splits into 8 sub-bands.
The correlated disorder potentials $w_k$ are generated 
iteratively site by site using a generalisation of the Gibbs
representation of Markov random chains \cite{JS87b}. 

The Gibbs distribution for the configuration of random variables 
$\mu$ is
\begin{equation}
P(\mu)=Z^{-1}\exp(-C\Theta(\mu)),
\end{equation}
where $\Theta(\mu)=-\sum_{<kl>} \sigma_k(\mu)
\sigma_l(\mu)$ is the Gibbs potential and $Z$ the partition
function. 
Here, $\sigma_k(\mu)=\mu_k \in [-1/2,1/2]$ 
is the value of the random variable on site $k$. 
For a given configuration $\mu^0$ the conditional 
probability that the new variable on the new site $n$ 
takes a value between $-1/2$ and $\Delta\le 1/2$ is
\begin{equation}
P(\mu_n<\Delta\,|\,\mu^0) = 
\frac{e^{\Delta C\mathit{\Sigma}_n}-e^{-1/2 C\mathit{\Sigma}_n}}
{2\sinh(1/2 C\mathit{\Sigma}_n)},
\end{equation}
where $\mathit{\Sigma}_n=\sum_{i\in N_n}\sigma_i$, and $N_n$ is the set of
neighbours of $n$ already determined. The correlation
parameter $C$ tunes the strength of the potential correlations. 
$C=0$ gives the uncorrelated case.

\begin{figure}
\centering
\includegraphics[width=7.5cm]{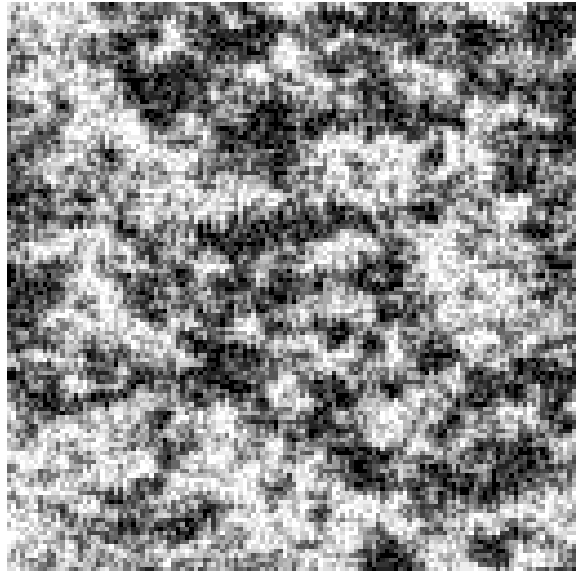}
\caption[]{%\small Fig.~\ref{examplecorr}
Gray-scale plot for a 128$\times$128 array of correlated disorder
potentials with correlation parameter $C=1.3$.}
\label{examplecorr}
\end{figure}

The correlated random numbers are then multiplied by the disorder strength 
$W$ which results in the set of disorder potentials $w_k$
used in Eqn.~(\ref{AH}).
An example of correlated potentials generated by this method with
$C=1.3$ is shown in Fig.~\ref{examplecorr} where a gray-scale plot 
of a 128$\times$128 array is displayed. White regions correspond to 
$w_k\approx +W$ and black areas to $w_k\approx -W$ with the gray
spots for values in between.

\begin{figure}
\centering
\includegraphics[width=7.5cm]{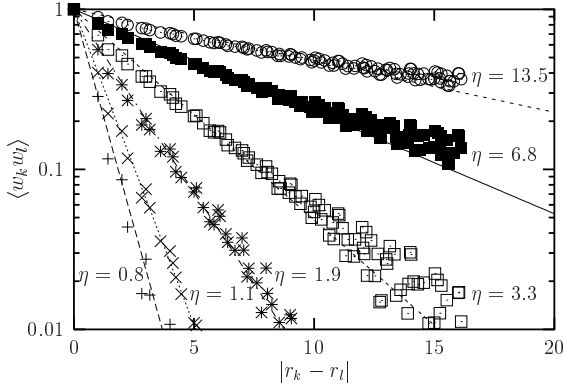}
\caption[]{%\small
The exponential decay of the correlations in the disorder potentials
for $C=0.5$ (+), 0.7 ($\times$), 1.0 ($\ast$), 1.3
(\raisebox{-.25mm}{$\Box$}), 1.7 (\rule{1.7mm}{1.7mm}), 
and 2.0 ({\large$\circ$}).}
\label{corrlength}
\end{figure}

To further characterise the disorder potentials, we have calculated the 
correlation function $K(\rho) = \langle w_k w_l\rangle_{\rho=|r_k-r_l|}$
averaged over all pairs of sites
which are a given distance $|r_k-r_l|$ apart,   
\begin{equation}
K(\rho)  =  
\frac{\sum_{i,j}^{N}w_i w_j\,\delta(|r_i-r_j|-\rho)}% 
{\left(\sum_i^{N} w_i^{2} \right)^{1/2}
\left(\sum_j^{N} w_j^{2} \right)^{1/2}}. 
\end{equation} 
For $C=0.5, 0.7, 1.0, 1.3, 1.7, \textnormal{\ and\ } 2.0$,
we find an exponential form, $K(\rho) \sim  \exp(-|r_k-r_l|/\eta)$. 
The spatial decay of the correlations is 
governed by the correlation length $\eta$ which in the range 
$0.7 \lesssim C \le 2.5$ can approximately be fitted by 
$\eta(C) \approx \exp(2C)$.
The results for the correlation function $K(\rho)$ and the
correlation length $\eta$ 
describing the decay of the potential correlations are shown 
in Fig.~\ref{corrlength}.

\section{Hall conductivity and localisation length}
The zero temperature Hall conductivity 
\setlength{\arraycolsep}{0.1cm}
\begin{eqnarray}
\sigma_{xy}(E,W) & = &
- \lim_{\varepsilon \rightarrow 0^+}\lim_{L^{\rule{0mm}{1.2mm}} 
\rightarrow \infty}
\frac{e^2}{h}\,\frac{2}{\Omega} {\rm Tr}
\bigg\{  
\\ 
\sum_{i}^{L} i \varepsilon (G_{ii}^{+}\nonumber &-&  
G_{ii}^{-}) x_i\, \hat{y}  - 2 \sum_{i,j}^{L} {\varepsilon}^2 
G_{ij}^{+} \, \hat{y} \, G_{ji}^{-} x_i   
\bigg\}
\end{eqnarray}
and the localisation length 
\begin{equation}
\lambda_M^{-1}(E,W)=-\lim_{L\to\infty}\frac{1}{2L}
\ln\textnormal{Tr}|G_{1,L}|^2
\end{equation}
are 
calculated numerically for large disordered two-dimensional systems 
of width $M$ and length $L$ using a recursive Green function method
developed previously \cite{Mac85,SKM85}. 
Depending on the
correlation length $\eta$ the width is varied between $M/a=32$ and $M/a=160$.
The length of the system necessary for convergence is typically
between $10^5$ and $10^7$ in units of the lattice constant $a$. 
In calculating the Hall conductivity one has strictly to observe the 
correct order of limits. 

\begin{figure}
\centering
\includegraphics[width=7.7cm]{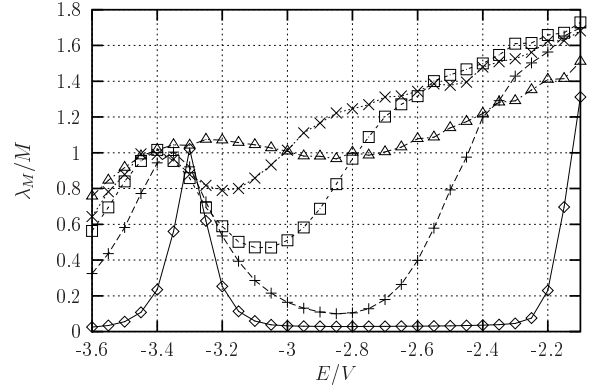}
\caption[]{%\small 
Normalised localisation length $\lambda_M(E,W)/M$  as a function of
energy $E$ for a system width $M/a=48$. The correlation parameter is $C=0.2$. 
The disorder strength is $W/V=0.5$ ($\Diamond$), 1.5 (+), 2.0
(\raisebox{-.25mm}{$\Box$}), 2.2 ($\times$), and 2.5 
({\scriptsize $\triangle$}).}
\label{W_M48C0p2_E}
\end{figure}

\section{Results and discussion}
The Hall conductivity was calculated near filling factor $\nu=2$ as
a function of increasing disorder $W$. For correlation parameters in the
range $0.0 < C < 1.0$ a continuous decay from $\sigma_{xy}=2e^2/h \to
\sigma_{xy}=0$ is observed. 
Although a direct transition is
seen in the Hall conductivity for these correlation parameters, a 
closer look at the corresponding localisation length indicates a 
totally different behaviour of the current carrying states. 
For $C=0.2$, which corresponds to a correlation length smaller than the
magnetic length, no floating up of the critical state of the lowest
Landau band can be observed. In fact, as in the
uncorrelated case it slightly moves downward in energy with increasing
disorder strength until it gets annihilated by the corresponding state
with negative Chern number moving faster downward from the band centre.
This is shown for a system of width $M/a=48$ in 
Fig.~\ref{W_M48C0p2_E} where the normalised
localisation length $\lambda_M/M$ is plotted versus energy in the 
filling factor range $0 \lesssim \nu \lesssim 1.5$ and
several disorder strength $W$. The transition to the insulator takes
place at $W_c/V\approx 2.5$ where the Chern and anti-Chern state
touch each other.  

\begin{figure}
\centering
\includegraphics[width=7.6cm]{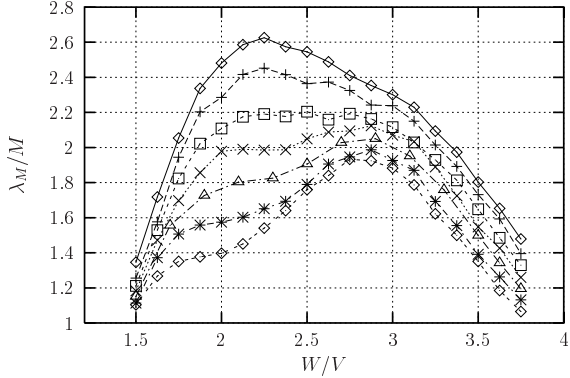}
\caption[]{%\small 
Normalised localisation length $\lambda_M(E,W)/M$ versus strength 
of the disorder potentials $W$. $C=1.0$ and $M=48\,a$. 
The energy parameter is $E/V=-1.2$ ($\Diamond$), $-1.3$ (+), $-1.4$
(\raisebox{-.25mm}{$\Box$}), $-1.5$ ($\times$), 
$-1.6$ ({\scriptsize $\triangle$}), 
$-1.7$ ($\ast$), and $-1.8$ ($\Diamond$ dashed line).}
\label{W_M48C1p0_E}
\end{figure}

A completely opposite behaviour is obtained for a larger correlation
parameter, $C=1.0$. This corresponds to a correlation decay length of
1.9\,$a$ which is almost twice as large as the magnetic length
$l_B/a=(2\pi\alpha)^{-1}$. While for $C=0.2$ the last current carrying
state disappeared at a disorder strength of about $W/V\approx 2.5$ two 
critical states are still observed at $W/V\approx 3.5$ and $C=1.0$.
This is shown in Fig.~\ref{W_M48C1p0_E} where the disorder dependence of 
the localization length for energies corresponding to filling factors 
$1.5 \lesssim \nu \lesssim 2.5$ is displayed. 
Fig.~\ref{W_M48C1p0_E} clearly reveals that there still exist two
distinct critical points which move to higher disorder values when the
filling factor is increased.  In other words, the current carrying
states of the two lowest Landau bands float up in energy and cross the
Fermi energy one by one. Within this energy (filling factor) range the
shift in disorder is weaker for the critical state of the lowest Landau 
band while it is stronger for the next critical state so that the two
states get closer with increasing filling factor. 

Since the two critical points are close in disorder the 
calculated Hall conductivity even for a system of width $M/a=160$ 
does not reflect the situation of two seperate transitions.
The reason for this behaviour is that the 
divergence of the localization length of the lowest 
and that of the next current carrying state 
overlap because for a given Fermi energy the two critical disorders 
are close (see abscissa in Fig~\ref{W_M48C1p0_E}) so that the Hall steps 
cannot be resolved. 
This overlap will eventually disappear in the thermodynamic limit and 
similarly will the Hall conductivity exhibit the missing Hall plateau at 
$\sigma_{xy}=e^2/h$. 
 
In conclusion, a lattice model with sufficient correlation in the disorder
potentials shows similar behaviour as the continuum model. Hence,
a lattice model can perfectly be in agreement with the proposed
levitation scenario \cite{Lau84,Khm84} and the predictions of the
global phase diagram \cite{KLZ92}. 

%\bibliographystyle{prsty}
%\bibliography{ludwig,papers2000_database}

\end{document}